\documentclass[useAMS,usenatbib]{mn2e}

\newcommand{\be}{\begin{equation}}
\newcommand{\ee}{\end{equation}} 
\newcommand{\bea}{\begin{eqnarray}}
\newcommand{\eea}{\end{eqnarray}}
\usepackage{graphicx}
\usepackage{tablefootnote}
\usepackage{color,soul}
\definecolor{lightgray}{gray}{0.85}
\sethlcolor{lightgray}
\title[The 2:3:6 QPO structure in GRS 1915+105 and cubic subharmonics]{The 2:3:6 QPO structure in GRS 1915+105 and cubic subharmonics
in the context of relativistic diskoseismology}

\author[M.~Ortega-Rodr\1guez et al.]
{M.~Ortega-Rodr\1guez$^{1,2,3}$\thanks{E-mail: manuel.ortega@ucr.ac.cr},
H.~Sol\1s-S\'anchez$^{1,3}$,
V.~L\'opez-Barquero$^{1,3}$,
\newauthor
B.~Matamoros-Alvarado$^{1,3}$ and
A.~Venegas-Li$^{1,3}$ \\
$^1$ Escuela de F\1sica \& 
Centro de Investigaciones Geof\1sicas, 
Universidad de Costa Rica, 11501-2060 San Jos\'e, Costa Rica \\
$^2$ Visiting Scholar, KIPAC, Stanford University, 
Stanford, CA 94305-4060 \\
$^3$ Instituto de F\1sica Te\'orica, 
1248-2050 San Jos\'e, Costa Rica}

\begin{document}

\maketitle

\begin{abstract}
We propose a simple toy model to explain the 2:3:6 QPO
structure in GRS 1915+105 and, more generally, the 2:3 QPO
structure in XTE J1550-564, GRO J1655-40 and H 1743-322.
The model exploits the onset of subharmonics in the context
of diskoseismology. 
We suggest that the observed frequencies may be the consequence of a 
resonance between a fundamental g-mode and
an unobservable p-wave. 
The results include the 
prediction that, 
as better data become available,  
a QPO with a frequency 
of twice the higher twin frequency 
and a large quality factor will be observed
in twin peak sources,
as it might already have been observed in the
especially active GRS 1915+105.
\end{abstract}


\section{INTRODUCTION}
 
In spite of active research,
the remarkable structure in the power spectra of
several X-ray binaries remains
a major puzzle for over a decade now.
Each one of the four black hole sources that show more
than one high frequency (40-450 Hz) quasi-periodic oscillation (HFQPO)
exhibits two of them in the 
`twin peaks' 2:3 ratio (see Table 1).

An understanding of HFQPOs may allow us to
obtain important information about the 
corresponding black hole's 
mass and spin, and the behavior of inner-disc accretion flows.

The physics of HFQPOs is not completely understood.
Since the observed 2:3 ratio suggests the presence of
non-linear physics, resonant models have been proposed.
Thus, 
starting with the pioneering work of 
\citet{Kluzniak},
explanations of the observed ratio have been sought 
by means of a parametric resonance 
among the dynamical (orbital and epicyclic)
frequencies (for more recent developments, see e.g.
\citealt{Torok1}; \citealt{Stuchlik}). 
Moreover, \citet{Kato} 
considers
long-wavelength
disc deformations which couple non-linearly to disc oscillations.
A detailed discussion of 
models can be found in \citet{Torok2}.

Even though
a considerable amount of research has thus already focused on 
non-linear resonances,
there is still room
for further exploitation of subharmonics,
especially given the fact that
other methods yield QPO frequencies which are too high to 
match observations \citep{Torok2}.

The main objective of this paper is thus to explore
the idea of subharmonics beyond
previous efforts.
Subharmonics, which have already been identified in stars
(e.g. in RV Tauri-type variables, \citealt{Pollard}),
have also been theoretically discussed in the context of
accretion discs by \citet{Kluzniak2} and \citet{Rebusco}, 
although both articles avoid details of disc models.

In this paper we would like to develop this
discussion by exploring the inclusion
of cubic (in addition to quadratic) 
subharmonics in the context of relativistic diskoseismology, 
the formalism of normal mode oscillations of thin accretion discs
(for a review, see \citealt{Wagoner1}).

According to diskoseismology, the 
observed oscillations 
in the outgoing X-ray radiation of 
systems such as GRS 1915+105
are due to normal modes of  
adiabatic hydrodynamic perturbations. These modes  
are the result of 
gravitational and pressure restoring forces in 
a geometrically thin, optically thick accretion disc
in the steep power-law state.

This interpretation is not only
corresponded      
observationally by narrow peaks in the 
power spectral density, but
some of these modes have been observed in 
hydrodynamic simulations as well \citep{Reynolds}.

Assuming this formalism is correct, 
we may use it to build
an exploratory study of non-linear effects.
If one thinks of diskoseismic modes as harmonic oscillators,
one might be able to model
non-linearities in the fluid equations
as non-linear terms added to a simple oscillatory system.

When devising a toy model
for the disc's complicated dynamics,
our aim was to propose
the simplest mathematical expression
which has solutions that include
quadratic and cubic subharmonics 
in a compact and observationally productive way.
It turns out that this can be achieved 
with a {\it single}
one-dimensional non-linear driven oscillator,
described by the following equation:
\be\label{main}
  \ddot{x} + \omega_0^2 x - \varepsilon x^2 - \delta x^3 = 
    B \cos \omega t 
\ee
(Landau \& Lifshitz 1976).
More complicated ODEs or
a system of coupled oscillators may be more 
realistic but at the expense of being less insightful.
Equation (\ref{main}) 
captures the essential
properties without unnecessarily obscuring the discussion. 
As we discuss below, 
this oscillator
features subharmonics that appear in a bifurcative way 
and
which are thus not obtainable by analytic continuation 
of linearized 
perturbation theory.

\begin{table} 
\centering
\caption{The frequencies of 2:3 QPO twin peaks in microquasars
and the corresponding black hole spin.
The values for the nondimensional black hole angular momentum parameter 
$a\equiv cJ/GM^2$
are averages of the different methods; numbers in
parentheses express the standard deviation when there is more than
one method.
References:
(1) \citealt{Wagoner2} (and references therein);
(2) \citealt{Torok2};
(3) \citealt{Remillard4};
(4) \citealt{McClintock};
(5) \citealt{Belloni2};
(6) \citealt{Remillard5};
(7) \citealt{Remillard3};
(8) \citealt{Remillard2};
(9) \citealt{Strohmayer};
(10) \citealt{Remillard1};  
(11) \citealt{Homan}.}
  \begin{tabular}{@{}lccl@{}}
 \hline
Source & Frequencies    &
Black Hole
& References
\\
 & (Hz) & Spin $(a)$ &  \\
\hline
GRS 1915+105  & 113$\pm$5    & 0.79 (19) & 1, 2, 3, 4, 5, 6 \\ 
        & 168$\pm$5    &     &   \\
XTE J1550-564 &  184$\pm$5   & 0.60 (28) & 1, 7 \\
        &  276$\pm$2   &     &    \\
GRO J1655-40  &  300$\pm$9   & 0.86 (14) & 1, 7, 8, 9\\
        &  450$\pm$5   &     &   \\
 H 1743-322   & 165$\pm$6    & 0.20 $\phantom{(14)}$  & 1, 10, 11  \\
        & 241$\pm$3    &     &    \\
\hline
\end{tabular}
\end{table}

The determination of the physical 
content of the terms
in (\ref{main}) constitutes the core of this paper, and 
this is the matter of Section 3.
One would like to know, for example, what fraction
of the hydrodynamic oscillation energy lies in 
non-linear interactions, and how would this information
be related to the peaks' properties in the power spectra. 
Before that, however, we will
describe the model in detail in Section 2,
while the main discussion, including numerical results and predictions, 
occupies Section 4. 

\section{THE MODEL}

\subsection{Background}

Modelling of non-linear oscillations in rotating stars
has been developed by 
\citet{Dyson}, \citet{Schutz1,Schutz2},
\citet{Kumar}, \citet{Wu}, \citet{Schenk},
and \citet{Arras}.
In this type of formalism, 
perturbation theory is used
to find the corrections to the linear r\'egime
in the form of non-linear couplings between 
(otherwise uncoupled) normal modes.
This type of formalism was applied by \citet{Horak}
to the case of slender tori as a rough approximation
to the oscillating region in an accretion disc.

One can thus calculate, in stars and discs, 
mode-coupling coefficients between three
or more modes.
These coefficients
provide important information about the system
(assuming non-linearities are mild),
such as selection rules for the participating modes and
relative strengths of the couplings,
in addition to slight changes in mode frequency (`detuning')
and stability considerations.

However, this approach has limitations
and fails once the amplitude of the oscillations 
becomes large enough
(even while still being in the perturbative r\'egime).
Then, matters can become intractable
as the appearance, via pitchfork bifurcation, of new forms of oscillation 
renders analytical approaches useless.

\citet{Vakakis} illustrates the concepts 
by means of a
toy system of two masses connected by non-linear springs.
Even though the system has only two degrees of freedom,
it can develop (under appropriate conditions) 
three forms of oscillation,
which he dubs {\it non-linear normal modes}. 

\subsection{Physical Interpretation}

Our toy model is a device that incorporates succinctly
the additional aforementioned modes,
and it is thus more than a
simplified visual depiction of the system.
We shall assume that equation (\ref{main}) represents
the dynamics of the QPOs
{\it after} it has reached a stationary state,
the (toy) variable $x$ being a measurement of fluid displacement.
Let us further assume that $\varepsilon$ and $\delta$
are small enough that perturbative considerations
make sense.

Assume that the oscillator 
described by the first two terms on 
the left-hand side of
equation 
(\ref{main}), $\ddot{x} + \omega_0^2 x$, 
represents a diskoseismic fundamental 
(axisymmetric) g-mode
(an inertial-gravitational oscillation), so we
set $\omega_0$ to be the g-mode frequency. Let us
call this frequency $\nu_2 \equiv \omega_0$.
Throughout this paper, we use the convenient notation
$\nu_n \equiv (n/2) \omega_0$.

Assume further that $\omega$ represents a higher 
frequency oscillation, associated with diskoseismic
axisymmetric
p-waves (inertial-acoustic oscillations).
Let us set $\omega = 3 \omega_0$ and thus 
call it $\nu_6 = \omega$. As shown in Fig.~1,
the conditions for the existence 
of axisymmetric g-modes and p-waves are that
the diskoseismic eigenfrequency $\sigma$ be 
smaller and greater, respectively, than the
radial epicyclic frequency $\kappa$.

\begin{figure}
\includegraphics[width=84mm]{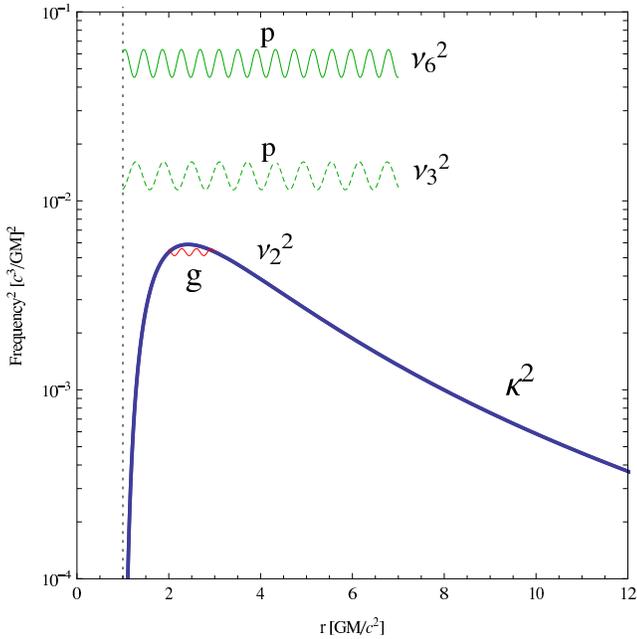}
\caption{The location of axisymmetric diskoseismic oscillation 
modes (wavy lines) and 
values of the square of their frequencies are plotted
as a function of the radius.
The $\nu_2$ frequency corresponds to the fundamental g-mode, while
$\nu_3$ and $\nu_6$ are p-waves.
Also shown are the square of $\kappa$ (the 
radial epicyclic frequency) for the case $a=1$, and
the position of the inner edge of the disc 
(dashed vertical line). The frequencies 
$\nu_2$, $\nu_3$ and $\nu_6$ are in a 2:3:6 relationship,
and thus $\nu_2$ and $\nu_3$ produce 
according to our model the twin peak pair.}
\label{figure}
\end{figure}

In the absence of the non-linear terms, the spectrum of 
the system described by equation
(\ref{main})
would contain $\omega_0$ and $\omega$, i.e. $\nu_2$ and $\nu_6$, 
and nothing else.
The presence of the term $\varepsilon x^2$ generates
a subharmonic of value $(3/2)\omega_0$
(i.e, half of $\omega$), which we call $\nu_3$,
while the term $\delta x^3$ is necessary
for the resonance between
the $\omega_0$ and the $\omega$ oscillators.
(For a theoretical discussion of subharmonics, see e.g.
\citealt{Jordan};
for experimental results, see e.g. \citealt{Linsay}.)

The frequencies $\nu_2$ and $\nu_3$ would constitute the observed
lower and higher QPO twin frequency pair.
But note that our model requires as well the existence of 
a $\nu_6$ QPO frequency (see Fig.~1).
Even though there is one notable instance of this 2:3:6 structure
in GRS 1915+105 
(\citealt{Remillard5}; 
\citealt{McClintock}),
this is rare and 
presumably the $\nu_6$ is in general either 
a transient phenomenon or
too weak to be observed, as non-linearities can render
subharmonics that are larger than the driving source.

We emphasize that equation (\ref{main}) represents an {\it effective} 
equation that describes the system in a formal way. 
The $\omega$ oscillator should not be regarded as
an `external' oscillator in a physical sense. 
Physically, it would
be more natural to think about the $\omega_0$ (g-mode) oscillator
as the one driving the motion. 
Thus, even though our model was conceived 
primarily to deal with the stationary state
of the disc, and not with its evolution,
the following sequential scenario 
suggests itself. 
Once a g-mode ($\nu_2$) and the p-wave ($\nu_6$) oscillations
are established via the cubic resonance,
then the quadratic resonance
generates a $\nu_3$ frequency.
In this way, the observed frequencies may be a consequence of a 
resonance between fundamental g-modes and
unobservable p-waves.
(A more comprehensive approach, involving 
all oscillators simultaneously, is described in Section 4.2.)

\subsection{Why Linear Diskoseismology is Insufficient}

A natural question to ask is just why higher frequency diskoseismic g-modes
cannot be used instead of p-waves as the $\omega$ resonator above.
After all, g-modes have frequencies in the right range,
and are the most robust modes, as they exist in the hottest part 
of the disc and lie away from the uncertain physics of the 
inner boundary \citep{Perez}. Modes with different angular mode number $m$
(such that oscillations are proportional to $e^{im\phi}$)
would seem to be good candidates.

For non-rotating black holes ($a = 0$), 
the frequencies corresponding to angular mode numbers
$m = 0$, 1, 2 are in a 1 : 3.5 : 6.4 relation.
For the $a = 0.5$ and
$a = 0.998$ cases, the relations
are 1 : 3.7 : 6.8 and 1 : 5.7 : 11.3,
respectively. 
(The dependence of the frequencies on the radial
and vertical mode numbers is very weak, 
changes amounting to a few percent or less.)

Thus, there are no modes close to three times the 
fundamental frequency.
In addition, the absence of negative $m$ g-modes 
causes the relevant non-linear couplings to vanish
due to  
selection rules of the form $\Sigma \, m_i = 0$ \citep{Arras}. 

\section{PHYSICAL MEANING OF THE PARAMETERS}

In order to further exploit the model,
we need to make a connection to more realistic 
(though not subharmonic producing) models.
In particular, we would like
to obtain values for $\varepsilon$ and $\delta$ in 
equation (\ref{main}) 
in terms of physical variables.

We will proceed thus to match our model
with the values obtained by \citet{Horak} for the
coupling coefficients of oscillations in slender tori
around of black holes.
This will allow us to obtain 
order-of-magnitude values
for our parameters.

We begin by reexpressing (\ref{main})
in the following form: (which is the one used by
\citealt{Horak})
\be\label{nond}
  \dot{y} + i \omega_0 y = i\omega_0 
(E y^2 + \Delta y^3) \, ,
\ee
in terms of the nondimensional quantities
\be
       y \equiv \frac{x}{A} \, ,
\quad
       E \equiv \frac{\varepsilon A}{3\omega_0^2} \, ,
\quad
       \Delta \equiv \frac{\delta A^2}{4\omega_0^2}  \, ,
\ee
where $A$ is the amplitude of the (toy) oscillation.
We have taken for now $B=0$, as
the interesting part is in the non-linear terms anyway,
and higher order terms $\propto \varepsilon^2$ have been dropped.

In this form, $y \sim 1$, 
while the conditions for the validity of the 
perturbative approach now read $|E| < 1$, $|\Delta| < 1$. 
The quantities $|E|$ and $|\Delta|$ have a simple physical meaning:
they represent the ratio of the non-linear interaction energy
to the energy of the linear mode (for the quadratic and cubic
terms, respectively).

Equations $\! (5), (6)$ and (8) in \citet{Horak} 
for the fluid displacement {\boldmath$\xi$} 
and the
coupling coefficients $\kappa$ and $b$ are:
(indices refer to modes)
\be
  \mbox{\boldmath$\xi$}(t,\mbox{\boldmath$x$}) = \sum c^*_A(t) \; 
  \mbox{\boldmath$\xi$}^*_A(\mbox{\boldmath$x$}) \, ,
\ee
\be\label{horak}
  \dot{c}_A + i \omega_A c_A = i b_A^{-1} {\cal{F}}_A^*  \, ,
\ee
\be
 {\cal{F}}_A = \sum \kappa_{ABC} \, c_B c_C
              + \sum \kappa_{ABCD} \, c_B c_C c_D + ... \, .
\ee
In this order-of-magnitude approach, 
we drop the indices in $c_A$ and $b_A$.
Since $|c| \sim 1$ for mild non-linearities,
we can now easily compare equations
(\ref{nond}) and (\ref{horak}), and identify $c$ with $y$.

This comparison can be readily performed 
once we have an order-of-magnitude estimate
for the (pressure- and gravity-dominated) couplings:
\be
   \kappa_{ABC}^{\rm (p)} \sim 
 \kappa_{ABC}^{\rm (g)} (R/h) \sim b \, \Omega (\xi/h) \, , 
\ee
where 
$\Omega$, $\xi$, $h$ and $R$ stand respectively
for the dynamical inverse time-scale and the 
displacement, disc thickness and disc radius length scales.
We have used $\xi/h \sim \Delta \rho/\rho$, the fractional
mass density, and
we have assumed that all the components
of $\xi$ have the same order-of-magnitude value.  

\section{DISCUSSION}

\subsection{General Results}

The considerations of the previous section allow
us to conclude, taking 
$\omega_0 \sim \Omega$, that
$|E| \sim \xi/h$ and $|\Delta| \sim E^2$.

Since we have that $|E|, |\Delta| < 1$ (our perturbative condition)
and that \cite{Nowak} found that $\xi/h \sim 1$,
we may conclude the following. 
$|E|$ has to be smaller than 1 yet not much smaller than 1
(meaning that our system is barely perturbative),
while $|\Delta|$ has to be smaller than $|E|$.
(It will be reassuring to obtain solutions with values consistent with this
reasoning below; see Table 2.)

Since $|\Delta|<|E|$,
it follows from (\ref{nond}) 
that the effects related to the 
quadratic term will be somewhat stronger than
those for the cubic one.
This means that the QPO peak amplitude should be 
higher for $\nu_3$ than for $\nu_2$,
which is
exactly what is observed. Remarkably,
this feature holds for {\it all} of the four sources
in Table 1. 
(For observational reviews, see \citealt{McClintock}, 
p.~157; \citealt{Belloni1}.)
The reasoning behind this statement becomes clearer
if one thinks of the non-linear terms as
driving forces.
 
Moreover, 
if one thinks of (\ref{nond}) as a spring (left-hand side)
with driving forces (right-hand side), and if one further
assumes there is a damping term 
$2 \beta \dot{x}$, with 
a quality factor $Q \equiv \omega_0/2\beta$,
then there will be, according to our model, 
an inverse relationship between the ratio 
$r \equiv$ amplitude of $\nu_3$/amplitude of $\nu_2$
(by ``amplitude'' we mean the standard {\sc psd} amplitude)
and $Q$.
This feature is also observed.
The variables
$r$ and $Q$
show a correlation of $-0.49$ when the value
of $Q$ for the lower twin frequency is used. 
(This correlation and the one in the next paragraph
have been calculated directly from the data in the 
references listed in Table 1.)

The plausibility of the model is further supported
by the following observation.
There exists a very strong correlation $(+0.91)$ between the amplitude
of the lower twin frequency and its $Q$,
while there is no significant correlation between the amplitude 
of the higher twin frequency and its $Q$.
This suggest that the two
QPO twin frequencies arise from
different physical mechanisms, and that 
the lower twin frequency
might play a more primary r\^ole
in the dynamics (e.g. that of being the driver of the 
QPO system), since it has a property that is degraded in the
other QPO frequency peak.
Such an interpretation is consistent with
the identification of the toy spring frequency
with the robust
diskoseismic fundamental g-mode.

\subsection{Parameter Determination} 

In order to obtain further results 
from our toy model, 
it is useful now to consider all of the 
relevant frequencies together. Thus,
we substitute
\be\label{all}
  x(t) =
C \cos(\omega t) + D \cos(\omega t/2) + F \cos(\omega t/3) + K
\ee
in (\ref{main}).
This procedure yields of course several cross terms in addition
to the purely quadratic and cubic subharmonic contributions.
By grouping the $\cos(\omega t)$, $\cos(\omega t/2)$ and 
$\cos(\omega t/3)$ terms,
it is thus possible to obtain a system of three coupled cubic
equations for $D/C$, $F/C$ and $K/C$ in terms of 
the parameters
$E$, $\Delta$ and $B/(C\omega^2)$,
all six quantities being nondimensional.

A numerical solution produces  
values for $(D/F)^2$, i.e.
the quantity we called $r$ in the previous section.
We assume that the non-linearities cause
locking of the phases of the terms in (\ref{all}),
as it is usual for non-linear systems
\citep{Pikovsky}.
There are only five different solutions for $(D/F)^2$
when scanning the parameter space
with the 65 000-point grid given by:
$0.1 < |E| < 0.5$ (step = 0.05),
$0.33 \, E^2 < |\Delta| < 3 \, E^2$ (step = 0.33),
$0.1 < |B/(8C\omega^2)| < 10$ (step = 0.1).

As Table 2 shows, most of the solutions have $(D/F)^2>1$.
The fourth solution is the closest to the observational
value of $\approx$ 2 (as can be appreciated directly from
the plots in the references of Table 1).
As expected, $|E|$ and $|\Delta|$ satisfy the properties
described at the beginning of Section 4.1.
The values of $B/(C\omega^2)$ are close to $-8$, its
value for the $\varepsilon = \delta = 0$ case.
 
Thus, even in the toy model approximation,
these numerical results 
yield a sensible outcome and have the expected 
parameter values.
They imply that
about half of the oscillation energy lies in 
non-linear interactions, especially those associated with the
quadratic subharmonic.

\begin{table}
\centering
\caption{The solution values of $(D/F)^2$
obtained numerically from equation~(\ref{all}), with
values of the parameters.
The quantity $(D/F)^2$ corresponds to the 
higher twin/lower twin 
{\sc psd} amplitude ratio.
$|E|$ is bigger than $|\Delta|$, showing dominance of
the quadratic subharmonic over the cubic one.}
\begin{tabular}{@{}cccc@{}}
\hline
$(D/F)^2$ & $E$ & $\Delta$ & $B/(C\omega^2)$ \\
\hline
          21\phantom{.00} & $\pm0.5$\phantom{0} & $\phantom{-}0.083$           &$-7.2$           \\
\phantom{0}2.4\phantom{0} & $\pm0.5$\phantom{0} & $\phantom{-}0.083$           &$-7.2$           \\
\phantom{0}4.6\phantom{0} & $\pm0.5$\phantom{0} & $-0.09$\phantom{0} &$-8$\phantom{.0} \\
\phantom{0}1.5\phantom{0} & $\pm0.45$           & $-0.073$           &$-8$\phantom{.0} \\
\phantom{0}0.16           & $\pm0.4$\phantom{0} & $-0.058$           &$-8$\phantom{.0} \\
\hline
\end{tabular}
\end{table}

\subsection{Predictions}

In the first place, the model predicts that, 
as better data and/or analysis become available,  
a QPO frequency equal to
$\nu_6$ will be revealed,
as it already appears to be present in GRS 1915+105. 
Furthermore, given that $\nu_6$ presumably forms before, 
and is the cause of, its
subharmonic $\nu_3$,
it is expectable
that $\nu_6$ will be less degraded,
i.e. have a larger value of $Q$,
than $\nu_3$ (as it already does 
for the case of GRS 1915+105). 
 
Secondly, 
QPO combination frequencies with values $\nu_3 \pm \nu_2$,
i.e. $\nu_1$ and $\nu_5$ in our notation,
may be observed, but {\it only}
when $\nu_2$ and $\nu_3$ are present at the same time. 
Even though the 1:2:3 harmonic observations from XTE J1550-564 
are favorable in this respect, there are currently 
not enough data
to state anything conclusive in this regard.

\subsection{Difficulties and Future Work}

A challenge of the model concerns 
explaining the
missing $\nu_4$,
given that 
the arguments applied to $\nu_6$ apply to $\nu_4$ as well,
i.e. why is there not a resonance between
$\nu_2$ and $\nu_4$ via the term $\varepsilon x^2$?

Table 3 summarizes the information of 
the different diskoseismic p-waves.
The size of each wave is estimated
by the diskoseismic radial wavelength 
\citep{Ortega}.

\begin{table}
\centering
\caption{The properties of theoretical
p-wave oscillations relevant to the discussion.
Wavelengths ($\lambda$) are obtained
from the diskoseismic formula
$\lambda^2 = h^2 ( 1 - {\kappa^2}/{\sigma^2})$,
where $h$ and $\kappa$ refer to to the disc's
thickness and the
radial epicyclic frequency, respectively.
Frequencies $\nu_3$, $\nu_4$ and $\nu_6$
are in a 3:4:6 relation.}
\begin{tabular}{@{}ccl@{}}
\hline
Frequency  & Size & \phantom{a}Observational status \\
($\sigma$) & & \\
\hline
    $\nu_6$ & $\lambda_6$ = 1.06 $h$ & 
\phantom{a}only observed in GRS 1915+105          \\ 
    $\nu_4$ & $\lambda_4$ = 1.15 $h$ & 
\phantom{a}not observed                           \\ 
    $\nu_3$ & $\lambda_3$ = 1.34 $h$ & 
\phantom{a}observed in all four sources \\
 & & \phantom{a}in Table 1\\ 
\hline
\end{tabular}
\end{table}

The physical behavior at the inner disc boundary is
probably the least understood aspect of the whole 
accretion disc system, given the sudden
domination of magnetic field and coronal effects there \citep{Hawley}. 
In particular,
it is not know whether there is wave leaking or reflection,
let alone phase change.

Assume, however, and for the sake of the present discussion, 
that there is enough of an impedance mismatch
at the inner boundary that the p-wave oscillations bounce back
without changing phase (assuming a free boundary condition).
Standing waves could then in principle be created for the waves in Table 3,
in what we may call ``diskoseismic semi-modes'' (since there is 
only one boundary, the inner one).
Alternatively, it may be the case that the non-linear
character of the system confines the oscillations 
away from the boundary as explained in \citet{Vakakis} 
for certain mechanical systems.

In either case,
since the distance $r(\kappa_{\rm max}) - r_i$
(the radius at which the radial epicyclic frequency is
maximum minus the radius of the inner disc boundary) 
is not
much larger than $h$, especially for fast spinning 
black holes \citep{Perez}, 
the diskoseismic p-waves will only travel a few
wavelengths before bouncing back and returning.

This sets up a scenario in which
$\nu_6$ and $\nu_3$ can build up but $\nu_4$ does not:
if the roundtrip distance,
which is 
given by twice $r(\kappa_{\rm max}) - r_i$,
equals e.g.
$5 \times \lambda_6 \approx 4 \times \lambda_3 \approx 4.6
\times \lambda_4$,
then there would be true $\nu_6$ and $\nu_3$ semi-modes,
while the necessary conditions for $\nu_4$ to exist 
will not be met,
as it would be out of phase.

If this model is on the right track, 
and thus a sizable fraction of the energy 
resides in non-linear interactions, then 
a more careful study of subharmonic dynamics is in order.
One might develop a full (coupled) model, 
or even
investigate the possibility of
solitonic confinement of the 
oscillations.

\section*{Acknowledgements}

This work was supported in part by grant 805-B2-176
of Vicerrector\'{\i}a de Investigaci\'on,
Universidad de Costa Rica. 
M.O.R. thanks Oficina de Asuntos Internacionales,
Universidad de Costa Rica,
for support during the Stanford visit.
We are grateful to diskoseismology collaborator Robert V. Wagoner,
Didier Barret, Miguel Araya, Andrey Herrera and an 
anonymous reviewer for helpful comments.

\end{document}